# Beyond Point-like Defects in Bulk Semiconductors: Junction Spectroscopy Techniques for Perovskite Solar Cells and 2D Materials

Ivana Capan

*Ruđer Bošković Institute, Bijenička 54, 10000 Zagreb, Croatia*

**Abstract**

Junction spectroscopy techniques (JSTs) are powerful tools for investigating electrically active defects in semiconductors. Originally developed to study point-like defects in bulk semiconductors, JSTs have since been extended to increasingly complex systems, providing valuable insights into defect energetics and interactions. This review paper outlines the fundamental principles of JSTs and critically examines their application to emerging materials, such as perovskite solar cells and two-dimensional (2D) materials. By highlighting both the capabilities and limitations of JSTs in these non-classical systems, the review demonstrates their continued relevance and important role in advancing next-generation semiconductor materials and devices.

## 1. Introduction

Junction spectroscopy techniques (JSTs), with deep-level transient spectroscopy (DLTS) as their primary workhorse, have been extensively used to investigate electrically active point-like defects in semiconductors since their introduction in 1974 by D. V. Lang [1]. Historically, these techniques have been applied predominantly to silicon and other group-IV semiconductors, enabling generations of researchers to develop a detailed understanding of defects, their physical properties, and their role in determining semiconductor performance [2,3,4,5]. By selectively probing electrically active defects, JSTs have also played a pivotal role in defect engineering, which has become essential for the development and optimization of modern electronic devices.

The rapid emergence of new semiconductor materials and device architectures has created an increasing demand for defect characterization in environments that are far more complex than the classical point-like defect systems for which junction spectroscopy techniques were originally developed. This evolution raises a fundamental question: to what extent, and under which conditions, can JSTs be reliably applied to systems in which the direct correspondence between, for example, DLTS signatures and isolated point-like defects, well established in silicon, is no longer straightforward?

In recent years, applications of JSTs have expanded significantly across a broad range of material systems. In many cases, the dominant electrically active defects are no longer isolated point defects, but are extended in nature (e.g., dislocations) [6,7], located at interfaces (Si/SiO$_2$ and 4H-SiC/SiO$_2$) [8,9,10], or embedded within nanostructured environments (e.g., Ge quantum dots in a SiO$_2$ matrix) [11,12]. Such systems challenge several of the fundamental assumptions underlying junction-based spectroscopy and complicate the interpretation of measured signals.

The primary aim of this review is to critically examine the applicability and limitations of junction spectroscopy techniques in complex semiconductor systems, with particular emphasis on perovskite solar cells and two-dimensional (2D) materials. Building on existing reviews that address the theoretical foundations of JSTs and their established role in the study of point-like defects in silicon, silicon carbide, and related materials, this work provides a focused overview of recent advances and emerging directions in applying JSTs to these non-classical material systems.

It should be emphasized that this review does not aim to revisit the theoretical foundations of junction spectroscopy techniques or to provide a detailed description of experimental setups, as comprehensive tutorials and reviews are already available in literature. Readers are particularly referred to the tutorial on JSTs by Peaker et al. [13] and the review on Laplace DLTS [14], which offer authoritative and thorough coverage of these topics.

Despite a growing number of individual studies applying junction spectroscopy techniques to halide perovskites [15, 16, 17] and 2D materials [18, 19], a critical review of the interpretation and limitations of JSTs in these material systems has not yet been published.

This paper is organized as follows. Section 2 provides an overview of the basic principles of junction spectroscopy techniques, with particular emphasis on deep-level transient spectroscopy. Section 3 introduces exemplary cases of point-like defects in bulk semiconductors and discusses sample preparation relevant for JST measurements. Sections 4 and 5 focus on the application of JSTs to perovskite solar cells and 2D materials, respectively. Finally, Section 6 presents the conclusions and outlook.

## 2. Junction Spectroscopy Techniques

In this section, a concise overview of the fundamental principles underlying JSTs is provided. Junction spectroscopy encompasses a range of measurement methods applied to semiconductor junctions, employing either electrical or electro-optical approaches [13,14,20]. The defining feature of these techniques is the use of a semiconductor junction to create a depletion region, which offers a distinct advantage over bulk characterization methods. In particular, the occupancy of defect-related energy levels within the bandgap can be controlled far more effectively in the depletion region than in the bulk material, enabling sensitive and selective probing of electrically active defects [13].

Defects that introduce energy levels deep within the bandgap are commonly referred to as deep-level defects.

More specifically, the deep-level defect processes relevant to semiconductor devices include: (i) generation of electrons and holes from deep levels, (ii) recombination of electrons and holes via deep levels, (iii) emission and capture of electrons by deep traps in n-type semiconductors, and (iv) emission and capture of holes by deep traps in p-type semiconductors. Figure 1 schematically summarizes the key defect-related processes that govern device behavior: carrier generation, recombination, and trapping through emission and capture.

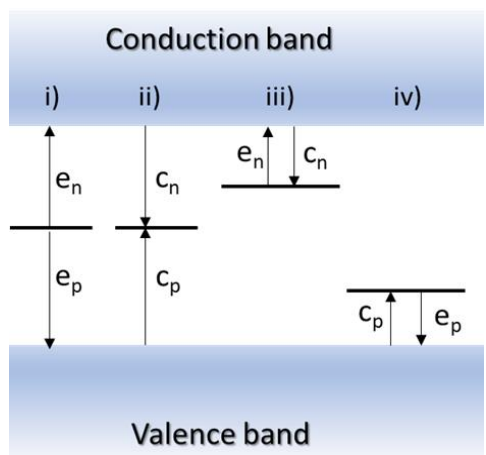

**Figure 1.** A schematic presentation for deep-level defect processes: i) generation of carriers (electrons and holes) from the deep level, ii) recombination of the carriers (electrons and holes) from the deep trap, iii) emission and capture of electrons from a deep level in a n-type semiconductor, and iv) emission and capture of holes in a p-type semiconductor. Here, $e_n$ is the electron emission rate, and $c_n$ is the electron capture rate, $e_p$ is the hole emission rate, and $c_p$ is the hole capture rate.

Among the various JST methods, DLTS is the most widely employed for studying deep-level defects. DLTS provides access to key defect parameters, including electron and hole emission activation energies, capture cross sections, and defect concentrations. Notably, DLTS can detect electrically active defects at concentrations as low as ~$10^{10}$ cm$^{-3}$ [13], making it a highly sensitive tool for defect characterization.

A typical DLTS measurement involves the repetitive filling and emptying of deep energy levels ($E_T$) within the depletion region of a Schottky barrier diode (SBD) using an applied bias pulse [20], as illustrated in Figure 2. For an n-type SBD, the diode is initially held under a reverse bias $V_R$ (Figure 2a), which is temporarily reduced to a filling pulse $V_P$ (Figure 2b). During the pulse, empty traps within the depletion region capture free carriers and become occupied. When the reverse bias $V_R$ is restored (Figure 2c), the presence of trapped charge reduces the net charge in the depletion region relative to its equilibrium value, giving rise to a measurable capacitance transient.

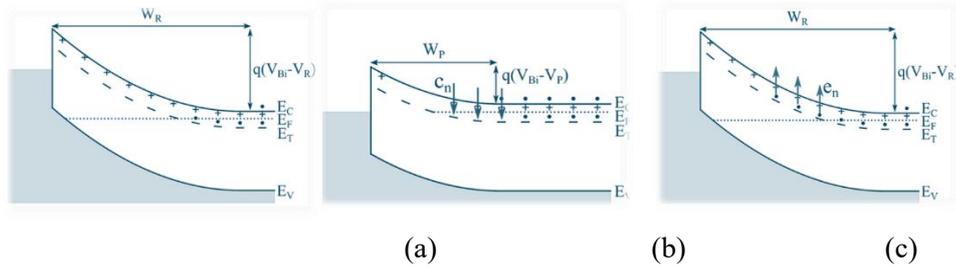

(a)　　　　　　　　(b)　　　　　　　　(c)

**Figure 2.** The basic principle of DLTS measurement: (a) equilibrium state with the applied reverse bias; (b) capture of the majority charge carriers in an n-type material while the pulse is applied $V_P$; (c) emission of the trapped charge carriers. Here, $W_R$ and $W_P$ are the depletion region widths for the applied $V_R$ and $V_p$, respectively, $e_n$ is the electron emission rate, and $c_n$ is the electron capture rate. Figure adapted from Ref. [20].

Following trap filling, the captured carriers are subsequently released via thermally activated emission, which proceeds exponentially in time. This thermal emptying of occupied traps is monitored by measuring the capacitance of a reverse-biased diode as a function of time after the application of the filling pulse (Figures 3a and 3b). The capacitance transient is sampled at two selected times, $t_1$ and $t_2$, and the DLTS signal is defined as the difference in capacitance, $\Delta C = C(t_1) - C(t_2)$. In a DLTS measurement, $\Delta C$ is recorded as a function of temperature, producing a characteristic DLTS spectrum as illustrated in Figure 3c.

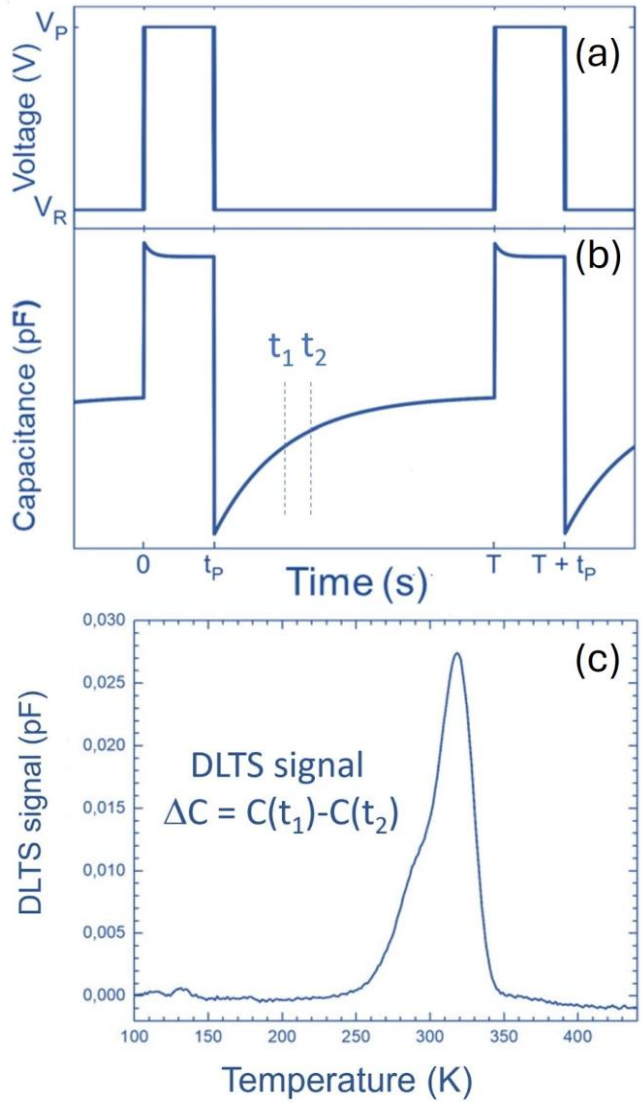

**Figure 3.** (a) The voltage changes, from $V_R$ to $V_P$ as a function of time; (b) The measured capacitance transient as a function of time, with marked selected times, $t_1$ and $t_2$. Here, $t_p$ is the filling pulse duration; (c) A typical DLTS spectrum; Figure adapted from Ref. [20].

A notable limitation of conventional DLTS is its relatively limited energy resolution, which makes it difficult to resolve closely spaced deep levels. This limitation is effectively overcome by Laplace DLTS, which improves energy resolution by approximately an order of magnitude, reaching the millielectronvolt (meV) range [14].

Laplace DLTS is an isothermal technique in which capacitance transients (measured following the same basic principles as in DLTS) are recorded and averaged at a fixed temperature. Instead of producing spectra as a function of temperature, Laplace DLTS yields a spectral representation of the processed capacitance signal as a function of the emission rate, obtained through an inverse Laplace transformation of the transient signal [13,14].

The ability of Laplace DLTS to provide additional physical insight has been demonstrated in numerous studies, including the identification of different charge states and configurations of defects such as carbon vacancy ($V_C$) and silicon vacancy ($V_{Si}$) in 4H-SiC [21,22,23]. Another illustrative example is the long-standing challenge of distinguishing between the $V_{Si}$ and the carbon interstitial ($C_i$) in 4H-SiC. Both defects can be introduced by irradiation, although $C_i$ is introduced by the low-energy electron irradiation (E<200 keV), as confirmed by Alfieri et al. [24] while $V_{Si}$ is introduced by MeV protons, neutrons and ion implantations [25,26,27].

Despite their different microscopic origins, both $C_i$ ($EH_1$) and $V_{Si}$ ($S_1$) introduce two deep-levels in the bandgap, with activation energies of 0.4 and 0.7 eV. As a result, DLTS spectra appear very similar and are impossible to distinguish, as shown in Figure 4a. In contrast, Laplace DLTS measurements, presented in Figures 4b and 4c, clearly resolve the differences between these defects. While the $EH_1$ level exhibits a single sharp Laplace DLTS peak, the $S_1$ level associated with $V_{Si}$ splits into two distinct Laplace peaks.

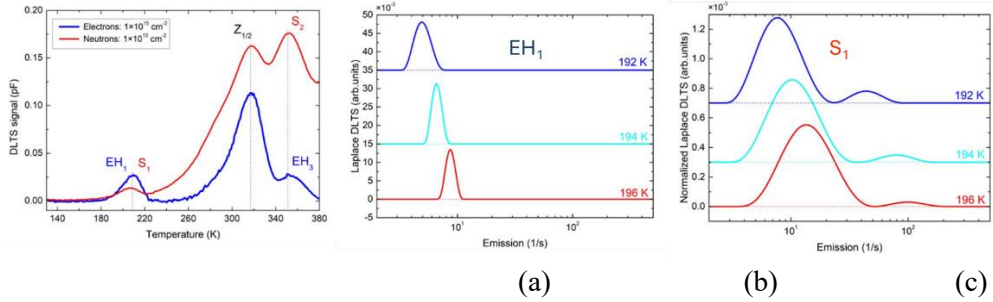

(a)　　　　　　　　(b)　　　　　　　　(c)

**Figure 4.** (a) DLTS spectra for low-energy electron (blue) and neutron (red) irradiated n-type 4H-SiC. (b) and (c) Laplace DLTS measurements at selected temperatures for the same samples. Measurement settings were $V_R = -10$ V, $V_P = 0$ V, and $t_P = 10$ ms. Figure taken from Ref.[26]

While DLTS and its variants, such as current DLTS (I-DLTS) and double DLTS (D-DLTS), have been predominantly used to investigate electrically active defects acting as majority carrier traps, minority carrier traps have been explored far less extensively. The fundamental principles of minority carrier transient spectroscopy (MCTS) were first introduced by Hamilton et al. [28] and subsequently refined by Brunwin et al. [29].

The key distinction between MCTS and DLTS lies in the method of carrier generation. In MCTS, minority carriers are generated optically using above-bandgap illumination [13,20]. An MCTS experiment involves the repetitive filling and emptying of deep levels through optical pulses with photon energy hν slightly exceeding the bandgap energy $E_g$, as illustrated in Figure 5. During the optical pulse, minority carriers are captured by deep levels (Figure 5a), while following the pulse these carriers are thermally emitted (Figure 5b).

In addition, sub-bandgap illumination may be employed to directly modify the occupancy of defect states according to their optical capture cross sections. This approach is commonly referred to as optical DLTS (O-DLTS) [13].

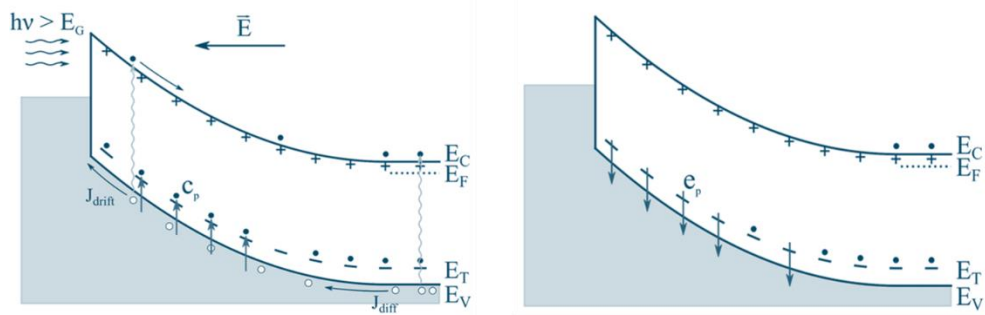

**Figure 5.** The basic principle of minority carrier transient spectroscopy (MCTS) measurement: (a) capture of the minority charge carriers from the valence band to the deep level ($E_T$) while optical pulses are applied; (b) emission of the minority charge carriers from the deep level after the optical pulse. Here,

$h\nu$ is the energy of the optical excitation, $J_{drift}$ and $J_{diff}$ are drift and diffusion current densities, respectively, $t_p$ is the optical pulse duration, $e_p$ is the hole emission rate, and $c_p$ is the hole capture rate. Figure adapted from Ref. [20].

Optical excitation in MCTS can be implemented either through a semi-transparent Schottky contact or via backside illumination of the sample. When the sample thickness exceeds the minority carrier diffusion length, thinning of the substrate is required to ensure efficient carrier collection within the depletion region [20].

It is also worth noting that minority carrier traps may be probed using forward-bias injection techniques. This approach, however, is subject to certain limitations, as discussed by Peaker et al. [13]. Nevertheless, in cases where the minority carrier capture cross section is significantly larger than that of the majority carrier, forward-bias-based methods can provide reliable and complementary information.

## 3. Point-like Defects in Bulk Semiconductors and Sample Preparation

This section briefly illustrates how the JSTs introduced in Section 2 have been successfully applied to the study of point-like defects in bulk semiconductors. Rather than providing an exhaustive overview, this section highlights a few representative examples and, most importantly, summarizes the typical sample preparation approaches such as vertically structured Schottky barrier diodes (SBDs) that have enabled the reliable and fruitful application of JSTs for decades. These well-established cases serve as a reference point for the more complex material systems discussed in the following sections.

Figure 6 summarizes the most common point-like defects in silicon carbide, namely the carbon vacancy ($V_C$), the silicon vacancy ($V_{Si}$), and the divacancy (VV). Among these, $V_C$ and $V_{Si}$ in 4H-SiC are among the most extensively studied defects. The $V_C$ is well known within the SiC community as a "lifetime killer," as it strongly reduces carrier lifetime and thereby affects device performance [30,31,32,33,34,35,36,37,38]. In contrast, $V_{Si}$ has attracted considerable interest for quantum technologies, ranging from solid-state qubits to single-photon emitters [22,25,26,27,39,40,41,42]. These defects were initially investigated using conventional DLTS. However, as already mentioned, significantly deeper insight has been achieved over the past decade through the application of Laplace DLTS [22,22,23,26].

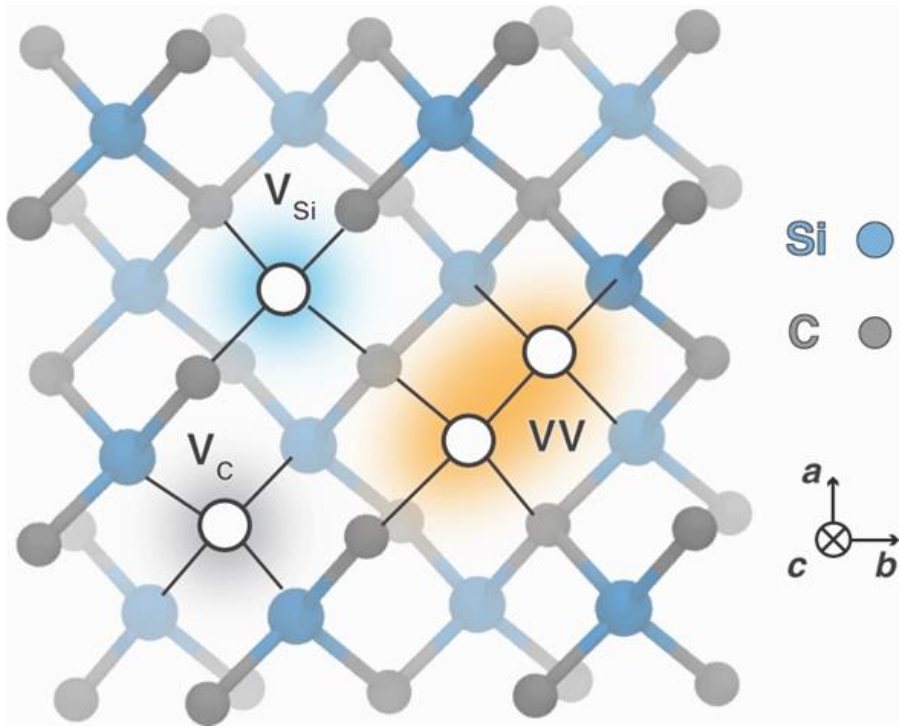

**Figure 6.** Typical point-like defects in SiC, include the silicon vacancy ($V_{Si}$), carbon vacancy ($V_C$), and divacancy (VV). Figure taken from Ref. [43].

A question, therefore, arises: how can the full potential of JSTs be exploited to obtain reliable defect information that enables a deeper understanding and improved control of semiconductor materials? For bulk semiconductors, the answer is relatively straightforward.

As implied by their name, JSTs require a semiconductor junction to establish a depletion region. Consequently, the most commonly employed junctions are p–n junctions and Schottky barrier diodes (SBDs). Among these, SBDs represent the simplest device architecture and have proven highly effective across a wide range of JST methods, including DLTS and MCTS.

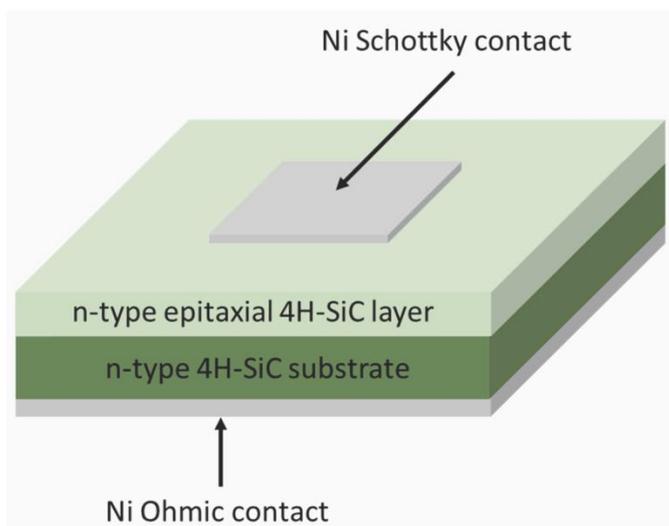

**Figure 7.** A typical SBD used for JSTs measurements. Figure taken from Ref. [44].

As shown schematically in Figure 7, the device consists of a Schottky contact on the front side and an ohmic contact on the back side of the semiconductor. The choice of Schottky metal depends primarily on its work function and the semiconductor type; for example, Ni is commonly used for n-type SiC, while Au is often employed for n-type Si [2,3,4,5,6,7,21,22,23,24,25,26]. Schottky contacts are typically formed by thermal evaporation through a shadow mask, defining contact areas on the order of a few mm² [44]. In contrast, the Ohmic contact is deposited on the entire backside of the wafer, usually without masking. For n-type 4H-SiC, Ni-based Ohmic contacts are commonly used, whereas Al is frequently employed for n-type Si [2,3,4,5,6,7,21,22,23,24,25,26].

The quality of both Schottky and Ohmic contacts is crucial for reliable JST measurements. Poor contact quality can lead to increased leakage currents, non-ideal capacitance behavior, and distorted transient signals. Therefore, contact performance is routinely evaluated prior to JST experiments using current–voltage (I–V) and capacitance–voltage (C–V) measurements to ensure proper rectifying behavior and stable depletion characteristics [13].

The procedure outlined above enables a straightforward and reliable analysis of point-like defects in bulk semiconductors. As illustrated in Figure 3, such defects produce well-defined capacitance transients, leading to sharp DLTS peaks and high-resolution L-DLTS signals, as shown in Figs. 3c and 4.

| Feature | Bulk Semiconductors (Si, Ge, SiC, GaN) | Perovskite Solar Cells | 2D Materials ($MoS_2$) |
|---|---|---|---|
| Dominant Defect Type | Point-like defects | Electronic and ionic defects | Point-like and interface-related defects |
| Junction | High-quality SBD or p-n junction | Complex heterojunction | Metal-insulator-semiconductor (MIS) |
| Transient | Exponential | Non-exponential (ionic-electronic mix) | Affected by interfaces and high contact resistance |

These characteristic features are summarized in Table 1, which also provides a comparative overview for perovskite solar cells and 2D materials. However, the favorable experimental conditions typically achieved in bulk semiconductors become considerably more difficult to realize in perovskite and 2D systems. The underlying reasons for these challenges are examined in the following sections.

## 4. Perovskite Solar Cells

Despite the remarkable rise of single-junction halide perovskite solar cells (PSCs) to power conversion efficiencies exceeding 27% [45], their performance and long-term stability remain strongly influenced by sub-gap defect states [46,47,48]. Accurate identification and quantification of these states are therefore essential for approaching the theoretical efficiency limit. JSTs represent powerful diagnostic tools for probing this hidden defect landscape [49, 50]. Figure 8 shows the most commonly observed defects in halide perovskite solar cells.

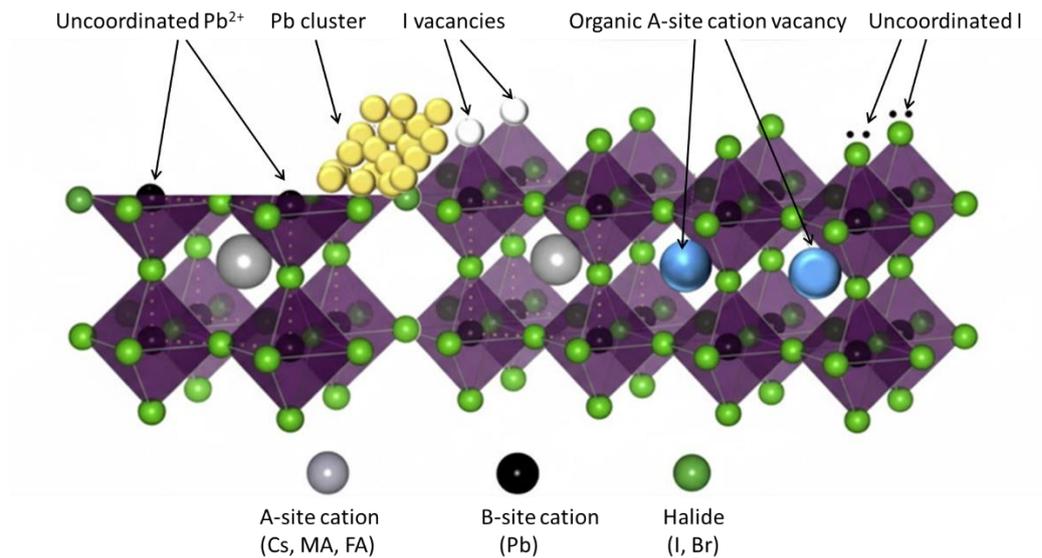

**Figure 8**. The most common defects in halide perovskites. Figure adapted from Ref.[51].

The defect environment in halide perovskites differs fundamentally from that of classical bulk semiconductors (Section 2). A unique challenge arises from the presence of mobile ionic defects [52]. Halide perovskites exhibit mixed ionic–electronic conductivity, where ionic species can migrate under external stimuli such as applied electric fields or illumination [53,54].

As illustrated in Figure 9, electronic and ionic processes follow distinct physical mechanisms:

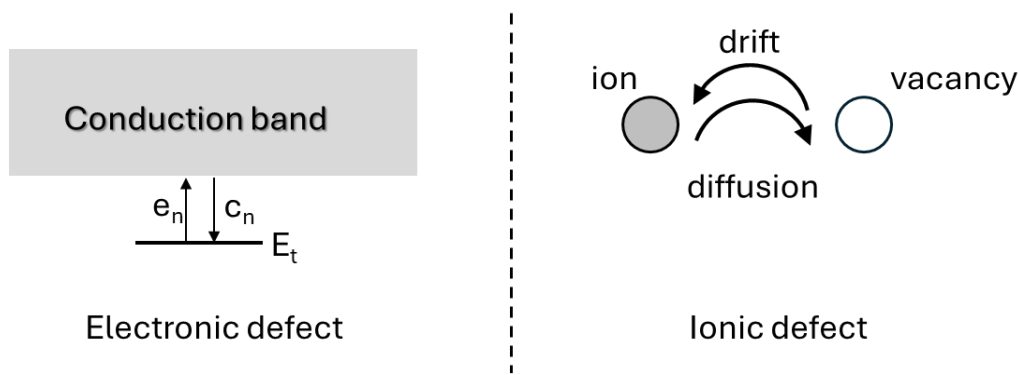

**Figure 9.** Electronic and ionic defect processes. For electronic defects, it is the emission and capture of electrons at the deep level within the bandgap ($E_t$). For ionic defects, it is the migration of mobile ions, drift and diffusion. Figure adapted from Ref.[55].

The main difference between electronic and ionic processes lies in their characteristic time scales. Electronic processes typically occur on microsecond-to-millisecond scales, whereas ionic processes often evolve over milliseconds to seconds [55,56,57,58].

Consequently, when applying DLTS to perovskites, the filling pulse duration ($t_p$) becomes a crucial experimental parameter. In conventional semiconductors, variation of $t_p$ is commonly used to distinguish point-like from extended defects through capture kinetics analysis [6]. In perovskites, however, extending the filling pulse into the seconds timescale is essential for isolating ionic contributions and avoiding misinterpretation of slow ionic relaxation as deep electronic trapping [55].

Reichert et al. [16] have applied DLTS to study defects in methylammonium lead triiodide (MAPbI3) solar cells in which defects were purposely introduced by fractionally changing the precursor stoichiometry. As seen in Figure 10, three temperature regions highlighted in gray correspond to three distinct defect-related DLTS peaks labelled as β, δ, and γ. The third peak, labeled γ, is not clearly resolved under all measurement conditions (for the rate window $t_2/t_1=10$) and becomes visible only for selected rate windows, as discussed in detail in the original work and its supplementary information [16]. It should be noted that the same group has performed comprehensive studies by applying a variety of JSTs techniques, such as reverse DLTS (the filling pulse is reversed, for example, the voltage pulse goes from 1 V to 0 V), Laplace DLTS, O-DLTS. I-DLTS, which have additionally confirmed their findings [50]. They have assigned all three observed defects to ionics defects: $V_{MA}^-$, $I_i^-$ and $MA_i^+$, respectively. These assignments partially overlap with those of Futcher et al. [59]. While also studying MAPbI3 perovskites, Futcher et al. [59] have observed three DLTS peaks. They have labeled them as A1, C1 and C2 and assigned A1 to $I_i^-$, and C1/C2 to $MA_i^+$.

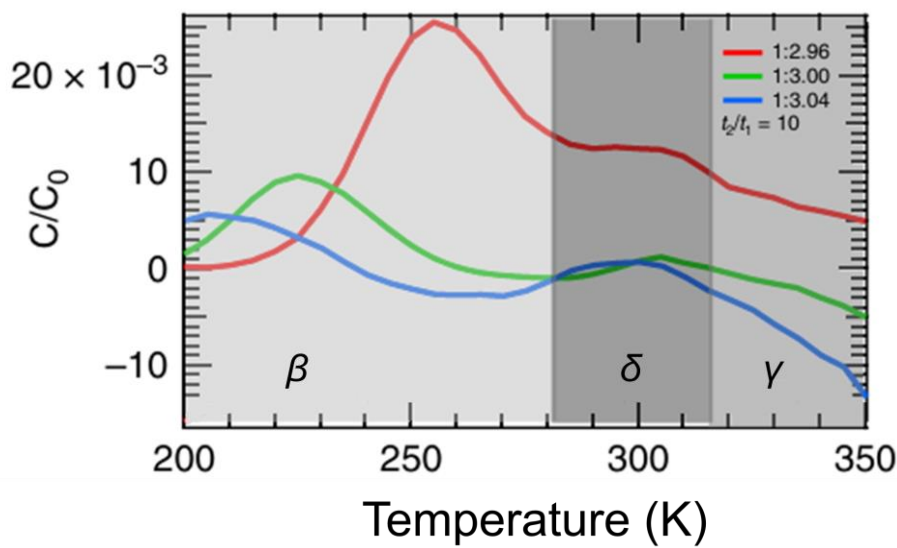

**Figure 10**. DLTS spectra for MAPbI3 perovskites with different precursor stoichiometry: 2.96 (red), 3.00 (green) and 3.04 (blue). Measurement settings were: $V_P$=1V, $V_R$=0V, $t_P$=100ms. Figure adapted from Ref. [16].

Another interesting feature observed in Figure 10 is the coexistence of positive (*β* and *δ*) and negative (*γ*) DLTS signals. This observation highlights an intrinsic advantage of DLTS: the ability to measure and distinguish majority and minority charge carriers. In bulk semiconductors, negative DLTS signals are typically associated with the minority carrier traps and can be probed under forward bias conditions [60] or by MCTS [61], while positive DLTS signals are associated with majority carrier traps [13].

In perovskites, interpretation requires extra caution due to mixed ionic–electronic transport. While the polarity of the DLTS signal yields deeper insight into the nature of defect states, it has introduced further ambiguity into defect labeling and identification, as demonstrated in the following examples.

Yang et al. [15] have applied DLTS to study formamidinium–lead–halide (FAPbI3) perovskite layers. In their study, they have shown that the introduction of additional iodide ions into the organic cation solution decreases the concentration of deep-level defects. They have identified three deep levels, labeled as A1, A2, and A3, with energies at 0.82, 0.78, and 0.46 eV below the conduction band. Notably, all observed DLTS peaks were positive, following the convention for electron traps in bulk semiconductors. Measurement setting were: $V_P=0$ V, $V_R= -0.7$ V, and the $t_p=100$ms. These defects are tentatively assigned to interstitial Pb ($Pb_i$) and antisite-related defects ($MA_I$, $Pb_I$, $I_{MA}$, and $I_{Pb}$).

Ren et al. [62] investigated perovskite solar cells based on mixed compositions of $FAPbI_3$ and $MAPbBr_3$ and observed three deep levels, labeled as H1, H2, and H3, with energies at 0.63, 0.68, and 0.77 eV above the valence band, respectively. It should be noted that DLTS peaks observed by Yang et al. [15] were all positive (as electron traps in bulk semiconductors), while DLTS peaks observed by Ren et al. [62] were all negative (as hole traps in bulk semiconductors). Figure 11a shows the DLTS spectrum for $(FAPbI_3)_{0.97}(MAPbBr_3)_{0.03}$ solar cell. H1, H2, and H3 are tentatively assigned to iodide vacancy ($V_I$), lead interstitial ($Pb_i$), and lead-iodide antisite ($Pb_I$) defects, respectively.

As stated in the previous section (Table 1), junction type in perovskite solar cells is more complex than a vertical SBD (Figure 5). The schematic illustration of the device structure of perovskite solar cell used by Ren et al. [62] is shown in Figure 11b.

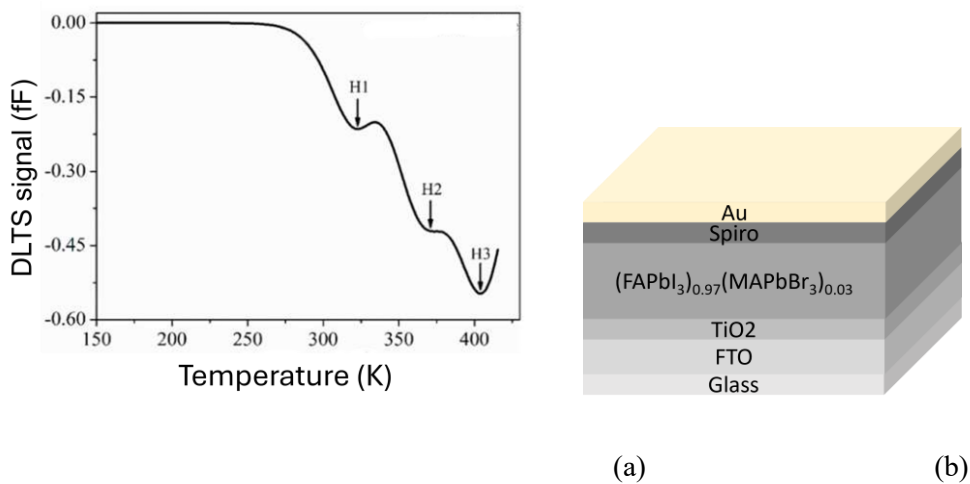

(a)          (b)

**Figure 11**. (a) DLTS spectrum for (FAPbI3)0.97(MAPbBr3)0.03 solar cell. Measurement settings were $V_P=0.6$V, $V_R=0.4$V, and tp=100ms. (b) Device structure used for DLTS measurements. Figure adapted from Ref. [62].

It is important to emphasize that while some studies report activation energies relative to $E_C$ or $E_V$, others report them simply as an absolute activation energy ($E_a$). Given that perovskite materials are frequently intrinsic or unintentionally doped, the standard convention of assigning trap types based on DLTS peak polarity can be ambiguous. In these non-classical systems, reporting the value as $E_a$ provides a more accurate and conservative representation of the experimental findings, avoiding potential misidentification of the defect's nature.

A summary of the defect fingerprints from the aforementioned studies is provided in Table 2. Notably, a significant discrepancy in defect identification remains, largely arising from contradictory peak polarities in the reported literature. Achieving a more reliable defect library for perovskite solar cells will require a deeper understanding of the interplay between electronic and ionic species. Furthermore, moving beyond a simple comparison of peak signs toward a rigorous analysis of transient behavior and consistent defect labeling will be essential for establishing a standardized consensus in the field.

**Table 2.** Electricaly active defects observed in perovskites by DLTS. Activation energies are reported as provided in original sources.

| Perovskite | Defect label | Activation energy (eV) | Identification | Reference |
|---|---|---|---|---|
| FAPbI$_3$ | A1 | $E_C - 0.82$ | Pb$_i$ and antisite defects | [15] |
| FAPbI$_3$ | A2 | $E_C - 0.78$ | Pb$_i$ and antisite defects | [15] |
| FAPbI$_3$ | A3 | $E_C - 0.46$ | Pb$_i$ and antisite defects | [15] |
| MAPbI$_3$ | β | 0.37* | $V_{MA}^-$, | [50] |
| MAPbI$_3$ | δ | 0.19* | $I_i^-$ | [50] |
| MAPbI$_3$ | γ | 0.37** | $MA_i^+$ | [50] |
| (FAPbI$_3$)$_{0.97}$(MAPbBr$_3$)$_{0.03}$ | H1 | $E_V - 0.63$ | V$_I$ | [62] |
| (FAPbI$_3$)$_{0.97}$(MAPbBr$_3$)$_{0.03}$ | H2 | $E_V - 0.68$ | Pb$_i$ | [62] |
| (FAPbI$_3$)$_{0.97}$(MAPbBr$_3$)$_{0.03}$ | H3 | $E_V - 0.77$ | Pb$_I$ antisite | [62] |
| MAPbI$_3$ | A1 | 0.29* | $I_i^{--}$ | [59] |
| MAPbI$_3$ | C1 | 0.90** | $MA_i^+$ | [59] |
| MAPbI$_3$ | C2 | 0.39** | $MA_i^+$ | [59] |

*migration of anion (charge type negative)

**migration of cation (charge type positive)

## 5. 2D Materials

This section explores the application of JSTs to 2D materials. Composed of layers only one to a few atoms thick, 2D materials exhibit strong in-plane covalent bonding and weak interlayer van der Waals interactions [63], giving rise to physical properties that differ substantially from their bulk counterparts [63,64]. While graphene initiated the rapid development of 2D material research, its lack of an intrinsic bandgap limits its utility in digital electronics [63,64,65]. Consequently, research has shifted toward semiconducting transition metal dichalcogenides (TMDs) with the general formula MX, where M denotes a transition metal (e.g., Mo or W) and X represents a chalcogen (S, Se, or Te) [65–67]. Among these, molybdenum disulfide (MoS$_2$) has emerged as a prototypical 2D semiconductor system [63,65].

In bulk semiconductors, DLTS relies on a well-defined depletion region formed within a vertical SBD [66]. However, when the semiconductor thickness approaches the monolayer limit, this classical concept is challenged. The reduction in dimensionality and the absence of a true "bulk" volume mean that the traditional space-charge region used to monitor carrier emission from deep-level defects may no longer be applicable [68,69]. Consequently, DLTS signals in monolayer systems can be dominated by interface-related defects and contact-related phenomena, potentially masking intrinsic defect fingerprints of the 2D semiconductor.

As noted earlier (Section 1), DLTS has been successfully applied to study interface-related defects [8,9,10]. In this context, the Fermi level pinning effect should also be considered. Fermi level pinning occurs at semiconductor surfaces or interfaces when the Fermi level becomes fixed at a specific energy within the bandgap, largely independent of the metal work function or semiconductor doping [8, 70]. This effect typically arises from a high density of interface-related defects (e.g., dangling bonds) that introduce localized energy levels in the bandgap and trap charge, thereby controlling the electrostatic potential and determining the effective barrier height. Temperature-dependent variations in Fermi level pinning further complicate DLTS (measurement as a function of temperature) data interpretation [8]. Being an isothermal technique Laplace DLTS therefore could be useful in bridging this challenge.

The first DLTS applications to $MoS_2$ were performed on thick, quasi-bulk layers, where conventional vertical SBDs could be fabricated [71], as illustrated in Figure 12. Sufficient thickness enabled the formation of a well-defined depletion region, analogous to bulk semiconductors.

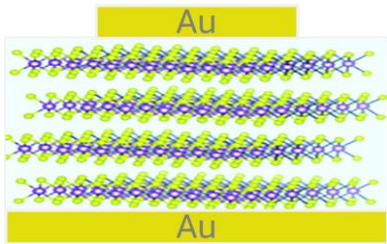

**Figure 12**. Vertical Schottky barrier diode fabricated on a thick $MoS_2$ layer.

Using this configuration, Kim et al. [70] applied DLTS and observed a deep-level defect with an activation energy of 0.35 eV (Figure 13). They have tentatively assigned this defect to sulfur vacancies (Vs). Their result was consistent with density functional theory (DFT) calculations for sulfur vacancy-related states.

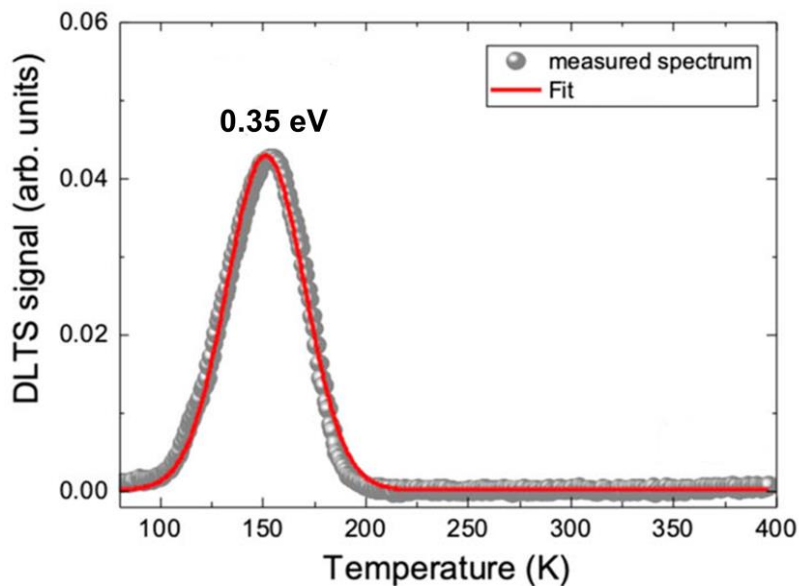

**Figure 13**. DLTS spectrum for bulk $MoS_2$. Measurement setting were: $V_R$=-1V, $V_p$=0V, $t_P$=1ms. Figure adapted Ref.[70].

As the thickness is reduced to a few-layer $MoS_2$, the vertical SBD geometry becomes increasingly problematic due to incomplete depletion and stronger contact effects. One strategy is the implementation of an inverse SBD configuration, where the junction is engineered to enhance depletion control in the thin layer. This approach was reported by Ci et al. [72]. Figure 14 shows a schematic of an asymmetric $MoS_2$ device, with a Schottky contact (Pt) at the bottom and an Ohmic contact (Ti) on top of a ~50 nm $MoS_2$ multilayer.

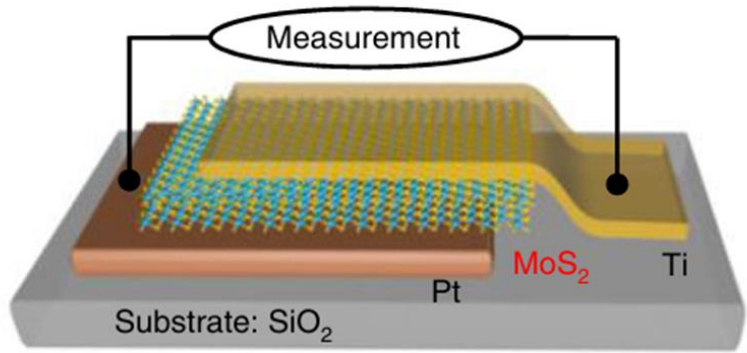

**Figure 14**. Inverse SBD configuration for multilayer MoS$_2$. Figure taken from Ref. [72].

DLTS measurements in this configuration revealed two distinct deep levels with activation energies of 0.27 and 0.40 eV. Measurement settings were: $V_R$=-0.5V, $V_P$=0V, $t_P$=1ms. The depleteion width was ~20nm.

Combining the DLTS results with first-principles calculations, the 0.27 eV defect was assigned to sulfur vacancy ($V_S$), while the 0.40 eV defect was attributed to a DX center, characterized by large lattice relaxation and metastable behavior, which is a significant finding for material stability. The DX center is a complex defect characterized by large lattice relaxation (LLR), where the defect atom undergoes a significant structural displacement upon capturing a carrier. This physical shift creates a large energy barrier for carrier capture and emission, often leading to phenomena such as persistent photoconductivity, where the material's conductivity remains altered long after an external stimulus is removed [72, 73] .

The most advanced demonstration of DLTS on a monolayer MoS$_2$ was reported by Zhao et al. [74]. They employed a different sample preparation approach, fabricating a metal–insulator–semiconductor (MIS) capacitor instead of SBD. Figure 15 shows the MIS device configuration used for DLTS measurements. In the monolayer limit, the traditional vertical depletion model is fundamentally altered. Because the semiconductor is only a few atoms thick, the electric field from the gate electrodes in the MIS configuration tends to modulate the carrier density across the entire monolayer rather than forming a conventional space-charge region [75].

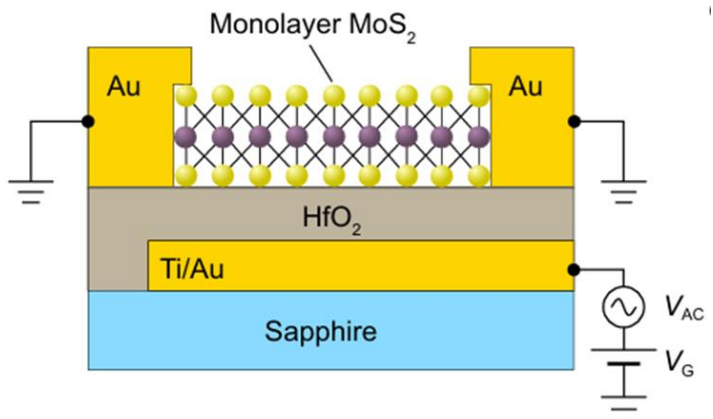

**Figure 15.** MIS device structure used for DLTS measuremens on a monolayer MoS2. Figure taken from Ref. [74].

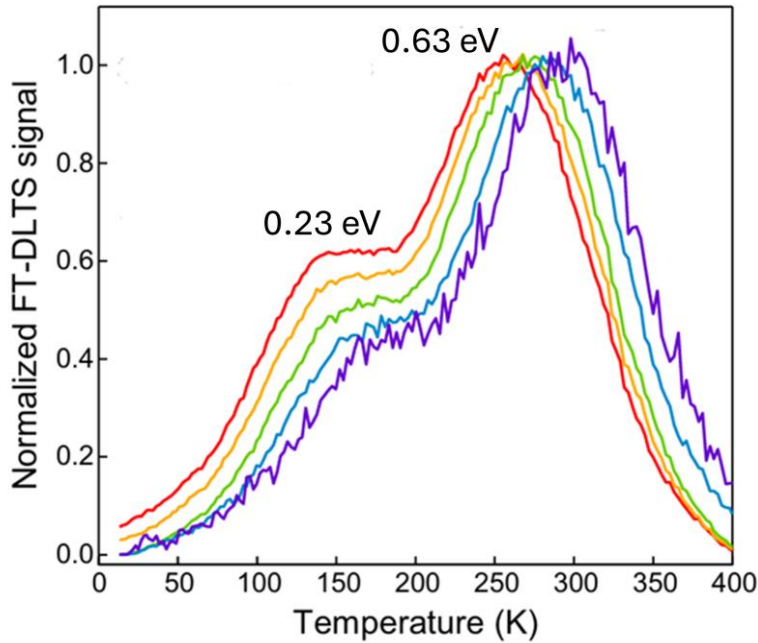

**Figure 16.** DLTS spectra for monolayer MoS2. Measurement settings were: $V_R$=0.75V, $V_P$=3.75V, $t_P$=100μs. Figure adapted from Ref.[74].

Two deep levels located at 0.23 eV and 0.63 eV below the conduction band were detected. The 0.23 eV was assigned to sulfur vacancies ($V_S$). Combining the DLTS results with scanning transmission electron microscopy (STEM) imaging and DFT calculations, authors have found out that neighbouring sulfur vacancy pairs are responsible for the second deep level (0.63eV) via hybridization of individual vacancy levels.

For additional information about DLTS measurements on MIS structure, and the role of electrical contacts readers are strongly encouraged to check work by Zhao et al. [74] including the supplementary material, as well as the comprehensive review paper on electrical contacts with 2D materials by Batool et al. [75].

The selected examples clearly demonstrate that, despite inherent challenges, DLTS provides invaluable insights into the electrically active defects in 2D materials. However, a direct comparison of activation energies across bulk-like, multilayer, and monolayer systems remains complex, primarily due to fundamental differences in device geometry, dimensionality, and interface effects. Notably, while additional JSTs such as Laplace DLTS, O-DLTS and I-DLTS have been successfully demonstrated in characterizing perovskite solar cells [50], their systematic application to 2D materials is still in its infancy, representing a promising frontier for future research.

## 6. Conclusions

Taken together, the selected examples discussed in this review highlight a clear message: JSTs have not merely kept pace with the evolution of semiconductor materials, they continue to contribute actively to our understanding of electrically active defects in increasingly complex systems. DLTS remains a central tool within the JST framework. Even in systems that differ fundamentally from the point-like defects in bulk semiconductors for which DLTS was originally developed, it continues to yield valuable physical insight. This sustained relevance reflects both the robustness of the underlying physical principles and the adaptability of the research community in extending DLTS to novel material classes.

At the same time, emerging semiconductor systems introduce new and substantial challenges. Mixed ionic–electronic transport, reduced dimensionality, complex heterostructures, and interface-dominated behavior complicate the interpretation of transient signals and demand greater methodological rigor. Extracting physically meaningful defect parameters in such environments will require carefully designed measurement protocols. The studies conducted by Reichart et al. [50] and Zhao et al. [74] have enabled further progress. Moreover, the integration of JSTs with computational modeling, data-driven analysis, and machine-learning (ML) tools offers a promising pathway toward deeper insight and more reliable defect identification [76,77,78]. The rapidly increasing number of studies employing ML-based defect identification indicates that this interdisciplinary direction will likely play a central role in future JST development.

As a conclusion, Table 3 summarizes the current status, main challenges, and possible future developments of JSTs across the material systems discussed in this review.

**Table 3**. The current status, main challenges and future directions of JSTs for bulk semiconductors, perovskite solar cells and 2D materials.

| Material System | Current Status of JST | Main challenge | Future Direction |
|---|---|---|---|
| Bulk semiconductors (Si, SiC, GaN etc.) | *Mature*. Comprehensively used for defect identification | Energy resolution. Overlapping peaks in Laplace DLTS | Automated defect-library matching using ML tools |
| Perovskite Solar Cells | *Experimental*. Emerging as a tool for exploring the defect landscape | Ionic-Electronic coupling. Distinguishing ionic from electronic defects. | DLTS *in situ* during light/heat cycles measurements |
| 2D Materials ($MoS_2$) | *Pioneering*. Mostly limited by sample preparation and contact issues | Depletion region and electrical contacts | Scanning-JST (combining DLTS with AFM/STM) |

**References**

1. Lang, D.V. Deep-level transient spectroscopy: A new method to characterize traps in semiconductors. J. Appl. Phys. 1974, 45, 3023–3032.
2. Svensson, B.G.; Mohadjeri, S.; Hallén, A.; Svensson, J.H.; Corbett, J.W. Divacancy acceptor levels in ion-irradiated silicon. Phys. Rev. B 1991, 43, 2292–2300.
3. Monakhov, E.V.; Wong-Leung, J.; Kuznetsov, A.Yu.; Jagadish, C.; Svensson, B.G. Electron emission characteristics of vacancy clusters and complex defects in irradiated silicon. Phys. Rev. B 2002, 65, 245201.
4. Svensson, B.G.; Jagadish, C.; Hallén, A.; Lalita, J.; Monakhov, E.V. Generation of vacancy-type point defects in single collision cascades during swift-ion bombardment of silicon. Phys. Rev. B 1997, 55, 10498.
5. Auret, F.D.; Peaker, A.R.; Markevich, V.P.; Dobaczewski, L.; Gwilliam, R.M. High-resolution DLTS of vacancy–donor pairs in P-, As- and Sb-doped silicon. Physica B: Condensed Matter 2006, 376–377, 73–76.
6. Capan, I.; Borjanović, V.; Pivac, B. Dislocation-related deep levels in carbon-rich p-type polycrystalline silicon. Solar Energy Materials and Solar Cells 2007, 91, 931.

7. Evans-Freeman, J.H.; Emiroglu, D.; Vernon-Parry, K.D.; Murphy, J.D.; Wilshaw, P.R. High-resolution deep level transient spectroscopy applied to extended defects in silicon. J. Phys.: Condens. Matter 2005, 17, S2219–S2227.
8. Dobaczewski, L.; Bernardini, S.; Kruszewski, P.; Hurley, P.K.; Markevich, V.P.; Hawkins, I.D.; Peaker, A.R. Energy state distributions of the Pb centers at the (100), (110), and (111) Si/SiO$_2$ interfaces investigated by Laplace deep level transient spectroscopy. Appl. Phys. Lett. 2008, 92, 242104.
9. Capan, I.; Pivac, B.; Slunjski, R. Electrical characterisation of Si–SiO$_2$ structures. Phys. Status Solidi C 2011, 8, 816–818.
10. Huang, W.; Dong, P.; Yang, N.; Ma, Y.; Xu, Q.; Fu, C.; Huang, M.; Li, Y.; Yang, Z.; Gong, M.; He, D.; He, Q. Effects of Nitrogen Passivation on the Capture Cross Section Energy Distribution of 4H-SiC/SiO$_2$ Interface Defects and the Temperature Dependences of Leakage Current. *Appl. Phys. Lett.* 2025, *127*, 122101.
11. Engström, O.; Kaniewska, M. Deep level transient spectroscopy in quantum dot characterization. Nanoscale Res. Lett. 2008, 3, 179–185.
12. Buljan, M.; Grenzer, J.; Holý, V.; Radić, N.; Mišić-Radić, T.; Levichev, S.; Bernstorff, S.; Pivac, B.; Capan, I. Structural and charge trapping properties of two bilayer films deposited on rippled substrate. Appl. Phys. Lett. 2010, 97, 163117.
13. Peaker, A.R.; Markevich, V.P.; Coutinho, J. Tutorial: Junction spectroscopy techniques and deep-level defects in semiconductors. J. Appl. Phys. 2018, 123, 161559.
14. Peaker, A.R.; Dobachewski, L. Laplace-transform deep-level spectroscopy: The technique and its applications to the study of point defects in semiconductors. J. Appl. Phys. 2004, 96, 4689.
15. Yang, W.S.; Park, B.-W.; Jung, E.H.; Jeon, N.J.; Kim, Y.C.; Lee, D.U.; Shin, S.S.; Seo, J.; Kim, E.K.; Noh, J.H.; Seok, S.I. Iodide management in formamidinium-lead-halide–based perovskite layers for efficient solar cells. Science 2017, 356, 1376–1379.
16. Reichert, S.; An, Q.; Woo, Y.-W.; Walsh, A.; Vaynzof, Y.; Deibel, C. Probing the ionic defect landscape in halide perovskite solar cells. Nat. Commun. 2020, 11, 6098.
17. Vasilev, A.A.; Saranin, D.S.; Gostishchev, P.A.; Didenko, S.I.; Polyakov, A.Y.; Di Carlo, A. Deep-level transient spectroscopy of the charged defects in p-i-n perovskite solar cells induced by light-soaking. *Opt. Mater. X* **2022**, *16*, 100218.
18. Ci, P.; Tian, X.; Kang, J.; Salazar, A.; Eriguchi, K.; Warkander, S.; Tang, K.; Liu, J.; Chen, Y.; Tongay, S.; Walukiewicz, W.; Miao, J.; Dubon, O.; Wu, J. Chemical trends of deep levels in van der Waals semiconductors. Nat. Commun. 2020, 11, 5373.
19. Zhao, Y.; Tripathi, M.; Čerņevičs, K.; Avsar, A.; Ji, H.G.; Gonzalez Marin, J.F.; Cheon, C.-Y.; Wang, Z.; Yazyev, O.V.; Kis, A. Electrical spectroscopy of defect states and their hybridization in monolayer MoS$_2$. Nat. Commun. 2023, 14, 44.
20. Capan, I.; Brodar, T. Majority and Minority Charge Carrier Traps in n-Type 4H-SiC Studied by Junction Spectroscopy Techniques. Electron. Mater. 2022, 3, 115-123.
21. Alfieri, G.; Kimoto, T. Resolving the EH6/7 level in 4H-SiC by Laplace-transform deep level transient spectroscopy. Appl. Phys. Lett. 2013, 102, 152108.
22. Bathen, M.E.; Galeckas, A.; Müting, J.; Ayedh, H.M.; Grossner, U.; Coutinho, J.; Frodason, Y.K.; Vines, L. Electrical charge state identification and control for the silicon vacancy in 4H-SiC. NPJ Quantum Inf. 2019, 5, 111.
23. Capan, I.; Brodar, T.; Pastuović, Z.; Siegele, R.; Ohshima, T.; Sato, S.I.; Makino, T.; Snoj, L.; Radulović, V.; Coutinho, J.; et al. Double negatively charged carbon vacancy at the h- and k-sites in 4H-SiC: Combined Laplace-DLTS and DFT study. J. Appl. Phys. 2018, 123, 161597.
24. Alfieri, G.; Mihaila, A. Deep level transient spectroscopy investigation of defects in semiconductors. J. Phys. Condens. Matter 2020, 32, 465703.
25. David, M.L.; Alfieri, G.; Monakhov, E.M.; Hallén, A.; Blanchard, C.; Svensson, B.G.; Barbot, J.F. Electrically active defects in irradiated 4H-SiC. J. Appl. Phys. 2004, 95, 4728–4733.

26. Knežević, T.; Brodar, T.; Radulović, V.; Snoj, L.; Makino, T.; Capan, I. Distinguishing the EH1 and S1 defects in n-type 4H-SiC by Laplace DLTS. Appl. Phys. Express 2022, 15, 101002
27. Brodar, T.; Capan, I.; Radulović, V.; Snoj, L.; Pastuović, Z.; Coutinho, J.; Ohshima, T. Laplace DLTS study of deep defects created in neutron-irradiated n-type 4H-SiC. Nucl. Instrum. Methods Phys. Res. Sect. B Beam Interact. Mater. Atoms. 2018, 437, 27–31.
28. Hamilton, B.; Peaker, A.R.; Wight, D.R. Deep-state-controlled minority-carrier lifetime in n-type gallium phosphide. *J. Appl. Phys.* 1979, 50, 6373–6385.
29. Brunwin, R.; Hamilton, B.; Jordan, P.; Peaker, A.R. Detection of minority-carrier traps using transient spectroscopy. Electron. Lett. 1979, 15, 349–350.
30. Kimoto, T.; Danno, K.; Suda, J. Lifetime-killing defects in 4H-SiC epilayers and lifetime control by low-energy electron irradiation. Phys. Status Solidi 2008, 245, 1327–1336.
31. Hazdra, P.; Popelka, S.; Schöner, A. Local Lifetime Control in 4H-SiC by Proton Irradiation. Mater. Sci. Forum 2018, 924, 436–439.
32. Galeckas, A.; Ayedh, H.M.; Bergman, J.P.; Svensson, B.G. Depth-Resolved Carrier Lifetime Measurements in 4H-SiC Epilayers Monitoring Carbon Vacancy Elimination. Mater. Sci. Forum 2017, 897, 258–261.
33. Saito, E.; Suda, J.; Kimoto, T. Control of carrier lifetime of thick n-type 4H-SiC epilayers by high-temperature Ar annealing. Appl. Phys. Express 2016, 9, 061303.
34. Miyazawa, T.; Tsuchida, H. Point defect reduction and carrier lifetime improvement of Si- and C-face 4H-SiC epilayers. J. Appl. Phys. 2013, 113, 083714.
35. Storasta, L.; Tsuchida, H. Reduction of Traps and Improvement of Carrier Lifetime in SiC Epilayer by Ion Implantation. Mater. Sci. Forum 2007, 556–557, 603–606.
36. Kushibe, M.; Nishio, J.; Iijima, R.; Miyasaka, A.; Asamizu, H.; Kitai, H.; Kosugi, R.; Harada, S.; Kojima, K. Carrier Lifetimes in 4H-SiC Epitaxial Layers on the C-Face Enhanced by Carbon Implantation. Mater. Sci. Forum 2018, 924, 432–435.
37. Hemmingsson, C.G.; Son, N.T.; Ellison, A.; Zhang, J.; Janzén, E. Negative- U centers in 4 H silicon carbide. Phys. Rev. B 1998, 58, R10119–R10122.
38. Son, N.T.; Trinh, X.T.; Løvlie, L.S.; Svensson, B.G.; Kawahara, K.; Suda, J.; Kimoto, T.; Umeda, T.; Isoya, J.; Makino, T.; et al. Negative-U System of Carbon Vacancy in 4H-SiC. Phys. Rev. Lett. 2012, 109, 187603.
39. Nagy, R.; Widmann, M.; Niethammer, M.; Dasari, D.B.R.; Gerhardt, I.; Soykal, Ö.O.; Radulaski, M.; Ohshima, T.; Vučić, J.; Son, N.T.; et al. Quantum Properties of Dichroic Silicon Vacancies in Silicon Carbide. Phys. Rev. Appl. 2018, 9, 034022.
40. Wang, J.; Zhou, Y.; Zhang, X.; Liu, F.; Li, Y.; Li, K.; Liu, Z.; Wang, G.; Gao, W. Efficient Generation of an Array of Single Silicon-Vacancy Defects in Silicon Carbide. Phys. Rev. Appl. 2017, 7, 064021.
41. Widmann, M.; Lee, S.Y.; Rendler, T.; Son, N.T.; Fedder, H.; Paik, S.; Yang, L.P.; Zhao, N.; Yang, S.; Booker, I.; et al. Coherent control of single spins in silicon carbide at room temperature. Nat. Mater. 2015, 14, 164–168.
42. Fuchs, F.; Stender, B.; Trupke, M.; Simin, D.; Pflaum, J.; Dyakonov, V.; Astakhov, G.V. Engineering near-infrared single-photon emitters with optically active spins in ultrapure silicon carbide. Nat. Commun. 2015, 6, 7578.
43. Lee, E.M.Y.; Yu, A.; de Pablo, J.J.; Galli, G. Stability and molecular pathways to the formation of spin defects in silicon carbide. Nat. Commun. 2021, 12, 6325.
44. Capan, I. 4H-SiC Schottky Barrier Diodes as Radiation Detectors: A Review. Electronics 2022, 11, 532.
45. National Renewable Energy Laboratory (NREL). Best Research-Cell Efficiency Chart. Available online: https://www.nrel.gov/pv/cell-efficiency.html (accessed on 19 February 2026).
46. Keeble, D.J.; Wiktor, J.; Pathak, S.K.; Phillips, L.J.; Dickmann, M.; Durose, K.; Snaith, H.J.; Egger, W. *Identification of lead vacancy defects in lead halide perovskites*. Nat. Commun. 2021, 12, 5566.

47. Wang, F.; Bai, S.; Tress, W.; Hagfeldt, A.; Gao, F. Defects engineering for high-performance perovskite solar cells. npj Flex. Electron. 2018, 2, 22.
48. Yuan, Y.; Yan, G.; Dreessen, C.; Hülsbeck, M.; Klingebiel, B.; Rudolph, T.; Ye, J.; Rau, U.; Kirchartz, T. Shallow defects and variable photoluminescence decay times up to 280 μs in triple-cation perovskites. *Nat. Mater.* 2024, *23*, 246–253.
49. Srivastava, S.; Ranjan, S.; Yadav, L.; Sharma, T.; Choudhary, S.; Agarwal, D.; Singh, A.; Satapathi, S.; Gupta, R.K.; Garg, A.; Nalwa, K.S. Advanced spectroscopic techniques for characterizing defects in perovskite solar cells. Commun. Mater. 2023, 4, 52.
50. Reichert, S.; Flemming, J.; An, Q.; Vaynzof, Y.; Pietschmann, J.-F.; Deibel, C. Ionic-Defect Distribution Revealed by Improved Evaluation of Deep-Level Transient Spectroscopy on Perovskite Solar Cells. Phys. Rev. Appl. 2020, 13, 034018.
51. Lu, H.; Liu, Y.; Ahlawat, P.; Mishra, A.; Tress, W.; Eickemeyer, F.T.; Yang, Y.; Fu, F.; Wang, Z.; Avalos, C.E.; et al. Compositional and Interface Engineering of Organic–Inorganic Lead Halide Perovskite Solar Cells. iScience 2020, 23, 101359.
52. Futscher, M.H.; Gangishetty, M.K.; Congreve, D.N.; Ehrler, B. Quantifying mobile ions and electronic defects in perovskite-based devices with temperature-dependent capacitance measurements: Frequency vs time domain. J. Chem. Phys. 2020, 152, 044202.
53. Eames, C.; Frost, J.M.; Barnes, P.R.F.; O'Regan, B.C.; Walsh, A.; Islam, M.S. Ionic transport in hybrid lead iodide perovskite solar cells. Nat. Commun. 2015, 6, 7497.
54. Futscher, M.H.; Milić, J.V. Mixed conductivity of hybrid halide perovskites: emerging opportunities and challenges. Front. Energy Res. 2021, 9, 629074.
55. Futscher, M. H.; Deibel, C. Defect spectroscopy in halide perovskites is dominated by ionic rather than electronic defects. ACS Energy Lett. 2022, 7, 140–144.
56. Pockett, A.; Eperon, G.E.; Sakai, N.; Snaith, H.J.; Peter, L.M.; Cameron, P.J. Microseconds, milliseconds and seconds: Deconvoluting the dynamic behaviour of planar perovskite solar cells. Phys. Chem. Chem. Phys. 2017, 19, 5959–5970.
57. Zhao, Y.; Zhu, K. Charge transport and recombination in perovskite ($CH_3NH_3$)$PbI_3$ sensitized $TiO_2$ solar cells. J. Phys. Chem. Lett. 2013, 4, 2880–2884.
58. Lou, F.; Yuan, S.; Wang, X.; Wang, H.-Y.; Wang, Y.; Qin, Y.; Ai, X.-C.; Zhang, J.-P. Distinguishing the migration time scale of ion species in perovskite solar cells. Chem. Phys. Lett. 2022, 796, 139570.
59. Futscher, M.H.; Lee, J.M.; McGovern, L.; Muscarella, L.A.; Wang, T.; Haider, M.I.; Fakharuddin, A.; Schmidt-Mende, L.; Ehrler, B. Quantification of ion migration in $CH_3NH_3PbI_3$ perovskite solar cells by transient capacitance measurements. *Mater. Horiz.* 2019, *6*, 1497–1503.
60. Kovačević, I.; Pivac, B.; Jaćimović, R.; Khan, M.K.; Markevich, V.P.; Peaker, A.R. Defects induced by irradiation with fast neutrons in n-type germanium. Mater. Sci. Semicond. Process. 2006, 9, 60–66.
61. Capan, I.; Brodar, T.; Yamazaki, Y.; Oki, Y.; Ohshima, T.; Chiba, Y.; Hijikata, Y.; Snoj, L.; Radulović, V. Influence of neutron radiation on majority and minority carrier traps in n-type 4H-SiC. Nucl. Instrum. Methods Phys. Res. B 2020, 478, 224–230.
62. Ren, X.; Zhang, B.; Zhang, L.; Wen, J.; Che, B.; Bai, D.; You, J.; Chen, T.; Liu, S. (F.); Deep-Level Transient Spectroscopy for Effective Passivator Selection in Perovskite Solar Cells to Attain High Efficiency over 23%. *ChemSusChem* 2021, *14*, 3182.
63. Nawaz, T. Graphene to Advanced $MoS_2$: A Review of Structure, Synthesis, and Optoelectronic Device Application. Crystals 2020, 10, 902.
64. Zhao, G.; Deng, H.; Tyree, N.; Guy, M.; Lisfi, A.; Peng, Q.; Nguyen, S.T.; Xiao, D.; Wang, K. Recent Progress on Irradiation-Induced Defect Engineering of Two-Dimensional 2D $MoS_2$ Few Layers. Appl. Sci. 2019, 9, 678.

65. Zhang, X.; Teng, S.Y.; Loy, A.C.M.; How, B.S.; Leong, W.D.; Tao, X. Transition Metal Dichalcogenides for the Application of Pollution Reduction: A Review. Nanomaterials 2020, 10, 1012.
66. Li, Y.; Tongay, S.; Yue, Q.; Kang, J.; Wu, J.; Li, J. Metal to semiconductor transition in metallic transition metal dichalcogenides. J. Appl. Phys. 2013, 114, 174307.
67. Edelberg, D.; Rhodes, D.; Kerelsky, A.; Kim, B.; Wang, J.; Zangiabadi, A.; Scurti, C.; Pasupathy, A.; Hone, J. Approaching the Intrinsic Limit in Transition Metal Diselenides via Point Defect Control. Nano Lett. 2019, 19, 4371–4379.
68. Celano, U.; Schmidt, D.; Beitia, C.; Orji, G.; Davydov, A.V.; Obeng, Y. Metrology for 2D materials: a perspective review from the international roadmap for devices and systems. Nanoscale Adv. 2024, 6, 2260–2269.
69. Kistanov, A.A. Characterization of Monovacancy Defects in Vanadium Diselenide Monolayer: A DFT Study. Appl. Sci. 2024, 14, 1205.
70. Kim, J.Y.; Gelczuk, Ł.; Polak, M.P.; Hlushchenko, D.; Morgan, D.; Kudrawiec, R.; Szlufarska, I. Experimental and Theoretical Studies of Native Deep-Level Defects in Transition Metal Dichalcogenides. npj 2D Mater. Appl. 2022, 6, 75.
71. Gelczuk, Ł.; Kopaczek, J.; Scharoch, P.; Komorowska, K.; Blei, M.; Tongay, S.; Kudrawiec, R. Probing Defects in $MoS_2$ Van der Waals Crystal through Deep-Level Transient Spectroscopy. Phys. Status Solidi RRL 2020, 14, 2000381.
72. Ci, P.; Tian, X.; Kang, J.; Salazar, A.; Eriguchi, K.; Warkander, S.; Tang, K.; Liu, J.; Chen, Y.; Tongay, S.; Walukiewicz, W.; Miao, J.; Dubon, O.; Wu, J. Chemical Trends of Deep Levels in van der Waals Semiconductors. Nat. Commun. 2020, 11, 5373.
73. Lang, D.V.; Logan, R.A. Large-Lattice-Relaxation Model for Persistent Photoconductivity in Compound Semiconductors. Phys. Rev. Lett. 1977, 39, 635–639.
74. Zhao, Y.; Tripathi, M.; Čerņevičs, K.; Avsar, A.; Ji, H.G.; Gonzalez Marin, J.F.; Cheon, C.-Y.; Wang, Z.; Yazyev, O.V.; Kis, A. Electrical Spectroscopy of Defect States and Their Hybridization in Monolayer $MoS_2$. Nat. Commun. 2023, 14, 44.
75. Batool, S.; Idrees, M.; Han, S.-T.; Roy, V.A.L.; Zhou, Y. Electrical Contacts With 2D Materials: Current Developments and Future Prospects. Small 2023, 19, 2206550.
76. Fried, H.P.; Barragan-Yani, D.; Libisch, F.; Wirtz, L. A Machine Learning Approach to Predict Tight-Binding Parameters for Point Defects via the Projected Density of States. *npj Comput. Mater.* **2025**, *11*, 1.
77. Tiwari, A.; Widodo; Krisnawati, D.I.; Kuo, T.-R. Physics-Informed Machine Learning for Accurate and Physically Valid Prediction of Semiconductor Defect Levels. Adv. Theory Simul. 2025, 9, 202501721.
78. Rahman, M.H.; Gollapalli, P.; Manganaris, P.; Yadav, S.K.; Pilania, G.; DeCost, B.; Choudhary, K.; Mannodi-Kanakkithodi, A. Accelerating Defect Predictions in Semiconductors Using Graph Neural Networks. APL Mach. Learn. 2024, 2, 016122.